\documentclass[aps,pre,groupedaddress,eqsecnum,twocolumn,amsmath,showpacs]{revtex4}
    \usepackage{bm}
    \usepackage{graphicx}
\newcommand\beq{\begin{equation}}
\newcommand\eeq{\end{equation}}
\newcommand\beqa{\begin{eqnarray}}
\newcommand\eeqa{\end{eqnarray}}
\newcommand{\nn}{\nonumber\\}

\newcommand{\py}{\text{PY}}
\newcommand{\mpy}{\text{mPY}}
\begin{document}
\title{Low-temperature
and high-temperature approximations for penetrable-sphere fluids.
Comparison with Monte Carlo simulations and integral equation
theories}
\author{Alexandr Malijevsk\'y}
\email{malijevsky@icpf.cas.cz} \affiliation{E. H\'ala Laboratory of
Thermodynamics, Academy of Science of the Czech Republic, Prague 6,
Czech Republic \\ Institute of Theoretical Physics, Faculty of
Mathematics and Physics, Charles University, Prague 8, Czech
Republic}
\author{Santos B. Yuste}
\email{santos@unex.es}
\homepage{http://www.unex.es/eweb/fisteor/santos/}
\author{Andr\'es Santos}
\email{andres@unex.es}
\homepage{http://www.unex.es/eweb/fisteor/andres/}
\affiliation{Departamento de F\'{\i}sica, Universidad de
Extremadura, E-06071 Badajoz, Spain}
\date{\today}

\begin{abstract}

The two-body interaction in dilute solutions of polymer chains in
good solvents can be modeled by means of effective bounded
potentials, the simplest of which being that of penetrable spheres
(PSs). In this paper we  construct two simple analytical theories
for the structural properties of PS fluids:  a low-temperature (LT)
approximation, that can be seen as an extension to PSs of the
well-known solution of the Percus--Yevick (PY) equation for hard
spheres, and a high-temperature (HT) approximation based on the
exact asymptotic behavior in the limit of infinite temperature.
Monte Carlo simulations for a wide range of temperatures and
densities are performed to assess the validity of both theories. It
is found that, despite their simplicity, the HT and LT
approximations exhibit a fair agreement with the simulation data
within their respective domains of applicability, so that they
complement each other. A comparison with numerical solutions of the
PY and the hypernetted-chain approximations is also carried out, the
latter showing a very good performance, except inside the core at
low temperatures.
\end{abstract}

\maketitle
\section{Introduction\label{sec1}}

Since  Wertheim and Thiele's analytical solution \cite{WT63} of the
Percus--Yevick (PY) integral equation for hard spheres (HSs), a
great deal of effort has been devoted to (approximately) describe
the equilibrium behavior of simple liquids. Potentials capturing
short-range harsh repulsion (with or without an attractive
contribution) have deserved particular interest. Recently, the
mainstream of attention has been transferred to more complex
liquids, such as colloidal dispersions. Interestingly enough,
coarse-graining techniques reveal that apparently toy interaction
models, some of them not diverging at zero separation, can serve as
excellent effective potentials in soft condensed matter.

While the effective interaction between two sterically stabilized
colloidal particles can be accurately modeled by the HS potential
\cite{L01}, the effective two-body interaction in other colloidal
systems can be much softer. For instance, the interaction potential
for star polymers in good solvents  diverges only logarithmically
for short distances \cite{L01,LLWAJAR98}. In the case of dilute
solutions of polymer chains in good solvents, the distance between
the centers of mass of two chains can be smaller than the sum of
their respective radii of gyration  \cite{L01}, so that the
effective two-body potential is {bounded}. This implies the
possibility of particles to form clusters, which makes the problem
greatly non-trivial. A simple class of bounded potentials is
described by the purely repulsive generalized exponential model
\cite{MFKN05,MGKNL06,LMGK07} parameterized by  an index $n$. In the
particular case $n=2$ one recovers the Gaussian core model, which
has  been widely studied
\cite{SS97,GL98,LLWL00,LBH00,LLWL01,FHL03,MKN06,Saija}. This model
exhibits re-entrant melting \cite{LLWL01} and the same is expected
if $n<2$ \cite{MFKN05,MGKNL06}. On the other hand,  a clustering
transition, which has been analyzed in detail for $n=4$
\cite{MGKNL06,LMGK07}, is predicted if $n>2$
\cite{LLWL01,MFKN05,MGKNL06}. In the limit $n\to\infty$, the
generalized exponential model reduces to the penetrable-sphere (PS)
model
\beq
\varphi(r)=\left\{
\begin{array}{ll}
\epsilon,& r<\sigma,\\
0,&r>\sigma.
\end{array}
\right.
\label{1}
\eeq
This model has been extensively investigated from different
perspectives
\cite{H57,GDL79,MW89,LLWL01,KGRCM94,LWL98,S99,FLL00,RSWL00,SF02,KS02,CG03,AS04,S05,S06,MS06,SM07}.
Density-functional theory \cite{LWL98,S99} predicts a freezing
transition to fcc solid phases with multiply occupied lattice sites.
The existence of clusters of overlapped particles (or ``clumps'') in
the PS crystal and  glass was already pointed out by Klein et al.\
\cite{KGRCM94}, who also performed Monte Carlo (MC) simulations on
the system. In the fluid phase, the standard integral equation
theories  are not  reliable in describing the structure of the PS
fluid inside the core  at low temperatures \cite{LWL98,FLL00}: the
number of overlapped pairs is overestimated by the hypernetted-chain
(HNC) theory, while it is strongly underestimated by the PY theory.
Other more sophisticated closures \cite{FLL00,CG03}, as well as
Rosenfeld's fundamental-measure theory \cite{RSWL00}, are able to
predict the correlations functions with a higher precision.

In any case,  the classical theories (PY and HNC), as well as the
alternative ones \cite{FLL00,CG03,RSWL00}, require numerical work to
get the correlation functions, such as the radial distribution
function $g(r)$ or, equivalently, the cavity (or background)
function $y(r)\equiv e^{\varphi(r)/k_BT} g(r)$, where $k_B$ is the
Boltzmann constant and $T$ is the temperature.

In general, the functions $g(r)$ and $y(r)$ convey the same physical
information. However,  the cavity function $y(r)$ is a much more
regular function than the radial distribution function $g(r)$. In
particular, if the interaction potential $\varphi(r)$ diverges in a
certain region (as happens for HSs with $r<\sigma$),  $g(r)$
vanishes in that region, while $y(r)$ keeps being well defined. In
that case, $y(r)$ provides more physical information than $g(r)$. In
this context, it is worth recalling the physical interpretation of
$y(r)$ as being proportional to the probability of accepting the
insertion of an additional test particle at a distance $r$ from a
given particle with which the test particle is assumed not to
interact \cite{LM84}.

Even in the one-dimensional (1D) case, the PS model is far from
trivial, since interactions are not restricted to nearest neighbors
and so its exact solution is not known \cite{MS06}. Useful
information, however, can be obtained in analytical or
semi-analytical form in some limiting cases. Thus, in the combined
high-temperature and high-density limit the PS model is amenable to
an  exact analytical treatment for any number of dimensions
\cite{AS04}. An interesting consequence of the boundedness of the PS
potential in the 1D case is the plausible existence of a
fluid-crystal phase transition \cite{AS04}, thus providing one of
the rare examples of phase transitions in 1D systems \cite{CS02}. In
the complementary low-density domain, the cavity function $y(r)$ has
been recently determined to second order in density at any
temperature \cite{SM07}. Comparison of the second-order contribution
to $y(r)$ with the corresponding PY and HNC predictions shows that
both are rather poor inside the core for low and moderate
temperatures (say $T^*\equiv k_BT/\epsilon\lesssim 1$) but the HNC
theory rapidly improves as the temperature increases. Finally, at
zero temperature ($T^*\to 0$), the PS model becomes identical with
the HS model. Although the correlation functions of HSs are not
exactly known (except in the 1D case), the PY theory is analytically
solvable for that interaction model \cite{WT63}, the corresponding
radial distribution function exhibiting a general good agreement
with computer simulations \cite{B74,HM86}. However, this good
agreement of the PY $g(r)$ for HSs can be misleading. As mentioned
before, $g(r)$ vanishes inside the core ($r<\sigma$) for HSs but the
cavity function $y(r)$ does not. It turns out that the PY $y(r)$ for
HSs is much worse for $r<\sigma$ than for $r>\sigma$ (even in the 1D
case \cite{MS06}), this deficiency being inherited by the PY
solution for PSs \cite{SM07}, in which case $g(r)\neq 0$ for
$r<\sigma$.

The main aim of this work is to construct two  simple analytical
theories for the structural properties of the three-dimensional PS
fluid. One of them is based on the exact solution of the PY equation
for HSs, modified inside the core. Since, as noted above, the HS
fluid is the zero-temperature limit of the PS fluid, this extension
of the PY solution is expected to be adequate for low and moderate
temperatures and so we will refer to it as the low-temperature (LT)
theory. The other approach, here referred to as the high-temperature
(HT) theory, is a simple extension to finite temperatures of the
exact asymptotic correlation functions in the limit $T^*\to\infty$.
Both approximations are compared with our own MC simulations as well
as with numerical solutions of the PY and HNC integral equations. It
is observed that the LT and HT theories complement quite well each
other, exhibiting a good agreement with the simulation data in their
respective domains of applicability, which are wider than what one
might have anticipated. It is also found that the HNC theory has an
excellent behavior, except inside the core at low temperatures,
while the PY results are always very poor inside the core.

In the next Section some basic properties of the PS fluid are
presented. Next, the zero-temperature limit is considered in Sec.\
\ref{sec3}, where special attention is paid to the PY solution for
HSs and an improved modified version of it in the overlapping region
is proposed. Our LT and HT theories are worked out in Secs.\
\ref{sec4} and \ref{sec5}, respectively. Section \ref{sec6} shows
the comparison with MC simulation data and with the PY and HNC
results for a wide range of representative states. Finally, the
paper is closed in Sec.\ \ref{sec7} with a summary and discussion of
the results.

\section{The PS model. Some basic results\label{sec2}}
Let us consider a three-dimensional fluid made of  particles
interacting via the pairwise PS potential \eqref{1}. The cavity
function $y(r)\equiv e^{\varphi(r)/k_BT} g(r)$ is a continuous
function of $r$, while the discontinuity of the potential at
$r=\sigma$ is transferred to the radial distribution function
$g(r)$:
\beq
g(r)=y(r)-x y(r)\Theta(1-r),
\label{2.1}
\eeq
where
\beq
x\equiv 1-e^{-1/T^*}
\label{3}
\eeq
is a temperature-dependent parameter bounded between $x=0$
($T^*\to\infty$) and $x=1$ ($T^*=0$), $\Theta(z)$ is the Heaviside
step function, and henceforth the distance is measured in units of
$\sigma$. The parameter $x$ represents the probability of rejecting
an overlap of two particles in an MC move. Some properties of the
Laplace transform $G(t)$ of $rg(r)$ are presented in Appendix
\ref{Laplace}.

In the limit of infinitely high temperatures at finite densities the
height of the potential barrier becomes negligible and so the system
behaves as an ideal gas. Apart from this trivial situation, the
exact form of $y(r)$ is not known, except (a) to second order in
density for any temperature \cite{SM07} and (b) in the
high-temperature and high-density limit \cite{AS04}.

\subsection{Low-density limit}
To first order in density, the cavity function  is exactly given by
\beq
y(r)=1+\frac{1}{2}\eta x^2
(2-r)^2(r+4)\Theta(2-r)+\mathcal{O}(\eta^2),
\label{17.2}
\eeq
where $\eta\equiv (\pi/6)\rho\sigma^3$, $\rho$ being the number
density. In the case of HSs, $\eta$ represents the packing fraction
and so here we will retain that name (although in the PS model there
is no a priori upper bound for it).
 {}From Eq.\ \eqref{17.2} one can easily get
\beq
y(1)=1+\frac{5}{2}\eta x^2+\mathcal{O}(\eta^2),\quad
y'(1)=-\frac{9}{2}\eta x^2+\mathcal{O}(\eta^2),
\label{II.1a}
\eeq
\beq
y(0)=1+8\eta x^2+\mathcal{O}(\eta^2),\quad  y'(0)=-6\eta
x^2+\mathcal{O}(\eta^2),
\label{II.1b}
\eeq
where the prime denotes a derivative with respect to $r$.  The
second-order contribution in the density expansion \eqref{17.2} is
also known \cite{SM07}, but it will not be needed here.
\subsection{High-temperature limit}
In the high-temperature and high-density limit ($T^*\to\infty$,
$\eta\to\infty$, $\eta x=\text{const}$), the cavity function has the
form \cite{AS04}
\beq
y(r)=1+ xw(r)+\mathcal{O}(x^2),
\label{2}
\eeq
where $w(r)$ is the inverse Fourier transform of
\beq
\widetilde{w}(k)=\frac{96\pi {\eta}x(k\cos k-\sin
k)^2}{k^3\left[k^3-24{\eta}x\left(k\cos k-\sin k\right)\right]}.
\label{4}
\eeq

The virial expansion of the function $w(r)$ is worked out in
Appendix \ref{app0}, where it is shown that the series converges for
$\eta x<\frac{1}{8}$. It is also proven in Appendix \ref{app0} that
$w(r)$ has a fourth-order discontinuity at $r=1$ and a discontinuity
of order $2(n-1)$ at $r=n\geq 2$.

The denominator in Eq.\ \eqref{4} has a real root if $\eta x\geq
\widehat{\eta}_0\simeq1.45$. This indicates the existence of a
spinodal transition (Kirkwood's instability) in the high-temperature
limit as $\eta x\to \widehat{\eta}_0$ from below. This instability
is preempted by the freezing transition, which can be estimated to
occur at $\eta x \approx 0.6$ \cite{AS04}.

\section{Zero-temperature limit\label{sec3}}
\subsection{Cavity function inside the core}
In the limit of zero temperature ($T^*\to 0$) the penetrable
particles become impenetrable, i.e., they become HSs. Therefore, in
that limit the radial distribution function $g(r)$ vanishes inside
the core ($r<1$). However, the cavity function is still well defined
in that region. In connection with this point, it might be useful to
recall two zero-separation theorems for HS fluids \cite{L95}:
\beq
\ln y(0)=\frac{1}{k_BT}\mu_{\text{ex}}=Z(\rho)-1+\int_0^\rho
d\rho'\frac{Z(\rho')-1}{\rho'},
\label{36}
\eeq
\beq
\frac{y'(0)}{y(0)}=-6\eta y(1).
\label{37}
\eeq
In Eq.\ (\ref{36}), $\mu_{\text{ex}}$ is the HS excess chemical
potential and $Z=1+4\eta y(1)$ is the HS compressibility factor.

\begin{figure}
\includegraphics[width=\columnwidth]{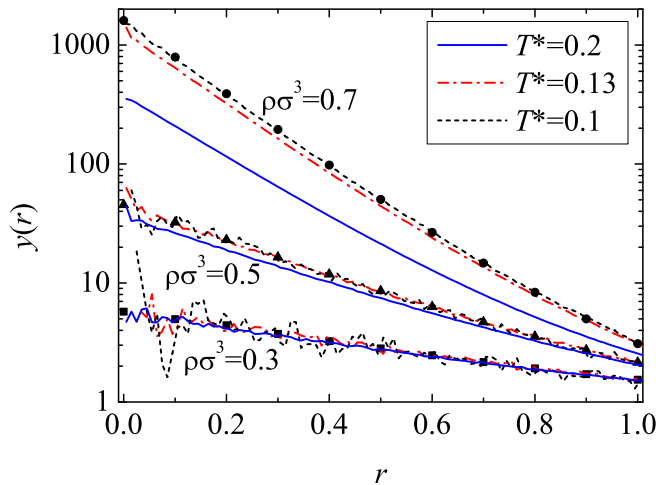}
\caption{(Color online) Plot of the cavity function $y(r)$ inside
the core ($r<1$) for densities $\rho\sigma^3=0.3$,
$\rho\sigma^3=0.5$, and $\rho\sigma^3=0.7$. The symbols are MC
simulation data  for HSs \protect\cite{LM84}. The lines are our MC
data for PSs at $T^*=0.2$ (solid lines), $T^*=0.13$ (dashed-dotted
lines), and $T^*=0.1$ (dashed lines).}
\label{y_HS}
\end{figure}
It is interesting to see how the cavity function for PSs converges
to that of HSs as the temperature decreases. Figure \ref{y_HS}
compares our MC simulation data obtained for PSs \cite{note} at
$T^*=0.2$ ($1-x\simeq 7\times 10^{-3}$), $T^*=0.13$ ($1-x\simeq
5\times 10^{-4}$), and $T^*=0.1$ ($1-x\simeq 5\times 10^{-5}$) with
the MC results obtained by Lab\'{\i}k and Malijevsk\'y for HSs
\cite{LM84}. Our PS data were obtained by measuring $g(r)$ directly
and then obtaining $y(r)$ inside the core as $y(r)=g(r)/(1-x)$. In
contrast, the HS data of Ref.\ \cite{LM84} were obtained by a
particle-insertion method.

We observe that for a density $\rho\sigma^3=0.3$ a temperature
$T^*=0.2$ is low enough to get results practically indistinguishable
from those of HSs ($T^*=0$). However, as the density increases one
needs to keep lowering the temperature  to get a good collapse.
Thus, for $\rho\sigma^3=0.5$  and $\rho\sigma^3=0.7$ the HS curves
are recovered at  $T^*=0.13$ and $T^*=0.1$, respectively. It is
interesting to note that, although at $T^*=0.2$ the probability of
accepting an attempted overlap between two PSs is only $1-x\simeq
7\times 10^{-3}$, a density $\rho=0.7$ is high enough to make the
corresponding cavity function  deviate strongly from that of HSs.
Figure \ref{y_HS} also shows that at low densities and temperatures
(e.g., $\rho\sigma^3=0.3$ and $T^*=0.1$) the number of overlaps is
very small and so the measured $y(r)$ presents large statistical
fluctuations.

\subsection{PY solution for HSs}
It is well known that the  radial distribution function $g(r)$ for
HSs is rather well represented by the PY solution
\cite{WT63,B74,HM86}. In that solution the auxiliary function $P(t)$
defined by Eq.\ \eqref{9} has the explicit rational form
\beq
 P(t)=-e^{-t} \frac{1+L_1 t}{1+S_1 t+S_2 t^2+S_3 t^3},
\label{13}
\eeq
where the coefficients can be univocally determined by application
of Eq.\ \eqref{17.2.1}. This yields \cite{YS91,HYS07}
\beq
L_1=\frac{1+\eta/2}{1+2\eta},
\label{43}
\eeq
\beq
S_1=L_1-1,
\label{44}
\eeq
\beq
S_2=\frac{1}{2}-L_1,
\label{45}
\eeq
\beq
S_3=\frac{6\eta L_1-1-2\eta}{12\eta}.
\label{46}
\eeq

Insertion of Eq.\ \eqref{13} into Eq.\ (\ref{9}) gives the Laplace
transform $G(t)$ for HSs. It can be expressed as
\beq
G(t)=\sum_{n=1}^\infty F_n(t) e^{-nt},
\label{47}
\eeq
where
\beq
F_n(t)\equiv -\frac{t}{12\eta}\left(\frac{1+L_1 t}{1+S_1 t+S_2
t^2+S_3 t^3}\right)^n.
\label{47.2}
\eeq
 The inverse Laplace transform of Eq.\ \eqref{47} yields
\beq
g(r)=\frac{1}{r}\sum_{n=1}^\infty f_n(r-n)\Theta(r-n),
\label{14hs}
\eeq
where $f_n(r)$ is the inverse Laplace transform of $F_n(t)$. In
particular,
\beq
y_\py(1)=\frac{1+\eta/2}{(1-\eta)^2},\quad
y'_\py(1)=-\frac{9\eta(1+\eta)}{2(1-\eta)^3},
\label{3.1}
\eeq
where henceforth  we use the label $\py$ in subscripts or
superscripts to indicate  explicit results derived from the PY
solution for HSs. The contact value $y_\py(1)$ yields the virial
route to the equation of state and from it one gets
\beq
\frac{1}{k_BT}\mu_{\text{ex}}^\py=2\ln(1-\eta)+\frac{2\eta(5-2\eta)}{(1-\eta)^2}.
\label{3.2}
\eeq

\begin{figure}
\includegraphics[width=\columnwidth]{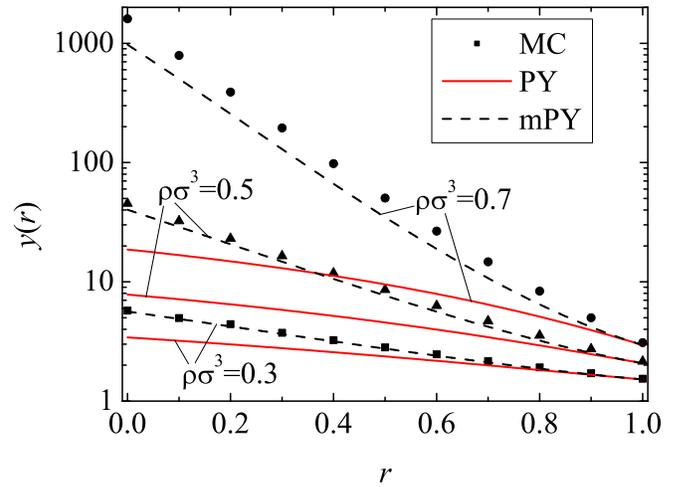}
\caption{(Color online) Plot of the HS cavity function $y(r)$ inside
the core ($r<1$) for densities $\rho\sigma^3=0.3$ (squares),
$\rho\sigma^3=0.5$ (triangles), and $\rho\sigma^3=0.7$ (circles).
The symbols are MC simulation data  \cite{LM84}, the solid lines are
the PY predictions, Eq.\ \protect\eqref{27}, and the dashed lines
are the modified PY predictions, Eq.\ \protect\eqref{m3.4}.}
\label{y_HS_PY}
\end{figure}

Equation (\ref{14hs}) gives reasonable predictions for the HS radial
distribution function \cite{HM86,B74}. However, the cavity function
in the region $r<1$, which  is given by
\beqa
y_\py(r)&=&\frac{1}{(1-\eta)^4}\Bigl[(1+2\eta)^2-6\eta(1+\eta/2)^2
r\nn &&+\frac{\eta}{2} (1+2\eta)^2r^3\Bigr],
\label{27}
\eeqa
strongly underestimates the correct values, even to second order in
density \cite{SM07}. {}From Eq.\ \eqref{27} one gets
\beq
y_\py(0)=\frac{(1+2\eta)^2}{(1-\eta)^4},\quad
\frac{y_\py'(0)}{y_\py(0)}=-6\eta\left(\frac{1+\eta/2}{1+2\eta}\right)^2.
\label{3.3}
\eeq
Equations (\ref{3.1}), (\ref{3.2}), and (\ref{3.3}) are not
consistent with the zero-separation theorems (\ref{36}) and
(\ref{37}). In fact, the values of $y(0)$ and $y'(0)$
thermodynamically consistent with $y_\py(1)$ are
\beq
y_\mpy(0)=(1-\eta)^2
\exp\left[\frac{2\eta(5-2\eta)}{(1-\eta)^2}\right],
\label{m3.3}
\eeq
\beq
\frac{y_\mpy'(0)}{y_\mpy(0)}=-6\eta\frac{1+\eta/2}{(1-\eta)^2},
\label{m3.3.2}
\eeq
where here the label $\mpy$ stands for ``modified'' PY values. One
can then modify the PY expression of $y(r)$ for $0\leq r\leq 1$ as
\beq
y_\mpy(r)=y_\py(r)e^{(r-1)^2(\lambda_0+\lambda_1 r)},
\label{m3.4}
\eeq
where $y_\mpy(1)=y_\py(1)$, $y_\mpy'(1)=y_\py'(1)$, and the
coefficients $\lambda_0$ and $\lambda_1$ are determined by imposing
consistency with Eqs.\ \eqref{m3.3} and \eqref{m3.3.2}, namely
\beq
\lambda_0=\ln\frac{y_\mpy(0)}{y_\py(0)},\quad
\lambda_1=2\lambda_0+\frac{y_\mpy'(0)}{y_\mpy(0)}-\frac{y_\py'(0)}{y_\py(0)}.
\label{m3.5}
\eeq
Of course, the modified form \eqref{m3.4} is not the only possible
one that can be proposed to satisfy the zero-separation theorems
with $y_\mpy(1)=y_\py(1)$ and $y_\mpy'(1)=y_\py'(1)$. The rational
behind Eq.\ \eqref{m3.4} is that, since the HS cavity function grows
so rapidly inside the core (see Fig.\ \ref{y_HS}), it is convenient
to add a polynomial correction to $\ln y_\py(r)$ rather than to
$y_\py(r)$.

Figure \ref{y_HS_PY} compares Eqs.\ \eqref{27} and \eqref{m3.4} with
MC data \cite{LM84}. It is quite apparent that the PY is very
inaccurate in the region $0\leq r\leq 1$. However, the simple
modified form \eqref{m3.4}, apart from incorporating the values of
$y(0)$ and $y'(0)$ consistent with the PY contact value $y_\py(1)$
via the zero-separation theorems, is quite satisfactory. On the
other hand, $y_\mpy(r)$ underestimates the true values, this effect
being especially visible at $\rho\sigma^3=0.7$. This is a
consequence of the fact that, as is well known, $y_\py(1)$ is lower
than the true contact value and this is transferred to $y_\mpy(0)$
 through Eq.\ \eqref{36}.

In order to extend Eq.\ \eqref{m3.3} to the PS case in Section
\ref{sec4}, it is convenient to express $y_\mpy(0)$ in terms of
$y_\py(1)$. To that end, we first rewrite Eq.\ \eqref{m3.3} in the
form
\beq
y_\mpy(0)=y_\py(1)e^{\alpha(\eta)\eta y_\py(1)},
\label{alpha}
\eeq
where
\beq
\alpha(\eta)\equiv
\frac{1}{1+\eta/2}\left[2(5-2\eta)+\frac{(1-\eta)^2}{\eta}\ln\frac{(1-\eta)^4}{1+\eta/2}\right].
\eeq
It turns out that $\alpha(\eta)$ depends very weakly on $\eta$,
deviating from $\alpha(0)=\frac{11}{2}$ less than $5\%$ for
$\eta<0.48$. Therefore, from a practical point of view we can
approximate \eqref{m3.3} or \eqref{alpha} by
\beq
y_\mpy(0)\simeq y_\py(1)e^{\frac{11}{2}\eta y_\py(1)}.
\label{11/2}
\eeq
In fact, we have checked that  Eqs.\ \eqref{m3.4} and \eqref{m3.5},
when implemented with Eq.\ \eqref{11/2} instead of Eq.\
\eqref{m3.3}, provides curves indistinguishable from the mPY curves
plotted in Fig.\ \ref{y_HS_PY}, except at the density
$\rho\sigma^3=0.7$, where Eq.\ \eqref{11/2} gives slightly better
results.

\section{The low-temperature (LT) approximation\label{sec4}}

Now we want to generalize to PSs the PY solution for HSs described
in Sec.\ \ref{sec3}, plus the modification \eqref{m3.4} for $r<1$,
so that  effects due to finite (but low) temperatures are
incorporated. The resulting approximation for $g(r)$  will be
essentially analytical, will reduce to the PY solution in the limit
$T^*\to 0$, and  will be better than the numerical solution of the
genuine PY integral equation for PS model, except at high
temperatures.

The steps are similar to the ones followed in the 1D case
\cite{MS06}. In a first stage, we keep the relationship (\ref{9})
between the Laplace transform $G(t)$ and the auxiliary function
$P(t)$ and propose an explicit form for the latter. In a second
stage, we modify the function $g(r)$ in the overlapping region  in
order to satisfy  approximate extensions of the zero-separation
theorems.

\subsection{First stage}
 Our  proposal for the function $P(t)$ defined by Eq.\ \eqref{9} is
\beq
 P(t)= \frac{L_0-1-e^{-t}(L_0+L_1 t)}{1+S_1 t+S_2 t^2+S_3 t^3},
\label{18}
\eeq
which reduces to the PY form \eqref{13} for HSs if $L_0=1$. However,
the presence of $L_0\neq 1$ for $T^*\neq 0$ (i.e., $x\neq 1$) is
necessary to satisfy the exact physical requirement \eqref{17bis.2}.
 Condition (\ref{17.2.1}) allows one to express the coefficients $L_1$, $S_1$, $S_2$, and
 $S_3$ in terms of $L_0$:
\beq
L_1=L_0\frac{1+\eta/2}{1+2\eta},
\label{43bis}
\eeq
\beq
S_1=L_1-L_0,
\label{44bis}
\eeq
\beq
S_2=\frac{1}{2}L_0-L_1,
\label{45bis}
\eeq
\beq
S_3=\frac{6\eta L_1-1-2\eta L_0}{12\eta}.
\label{46bis}
\eeq

Inserting (\ref{18}) into (\ref{9}) and expanding in powers of
$e^{-t}$, one gets
\beq
G(t)=\sum_{n=0}^\infty F_n(t) e^{-nt},
\label{47bis}
\eeq
where
\beq
F_0(t)=\frac{t}{12\eta}\frac{L_0-1}{L_0+S_1 t+S_2 t^2+S_3 t^3},
\label{48}
\eeq
\beqa
F_n(t)&=&-\frac{t}{12\eta}\frac{\left(1+S_1 t+S_2 t^2+S_3 t^3\right)
\left(L_0+L_1 t\right)^n}{\left(L_0+S_1 t+S_2 t^2+S_3
t^3\right)^{n+1}},\nn && n\geq 1.
\label{49}
\eeqa
 Therefore,
\beq
g(r)=\frac{1}{r}\sum_{n=0}^\infty f_n(r-n)\Theta(r-n),
\label{14}
\eeq
where $f_n(r)$ is the inverse Laplace transform of $F_n(t)$.
Equations \eqref{47bis} and \eqref{14} are formally analogous to
Eqs.\ \eqref{47} and \eqref{14hs}, respectively.  However,  some
important differences can be observed: Eqs.\ \eqref{47bis} and
\eqref{14} include extra contributions associated with $n=0$, Eq.\
\eqref{49} differs from Eq.\ \eqref{47.2}, and Eqs.\
\eqref{43bis}--\eqref{46bis} differ from Eqs.\
\eqref{43}--\eqref{46}. Obviously, Eqs.\ \eqref{43bis}--\eqref{14}
reduce to Eqs.\ \eqref{43}--\eqref{14hs} if $L_0=1$. Since $L_0\neq
1$ for PSs, the basic difference with respect to the HS case is that
now one has $g(r)=f_0(r)/r\neq 0$ for $r<1$. More specifically,
application of the residue theorem to get the inverse Laplace
transform of \eqref{48} yields
\beq
f_0(r)=
\frac{1}{12\eta}\sum_{i=1}^3z_i\frac{L_0-1}{S_1+2S_2z_i+3S_3z_i^2}e^{z_ir},
\label{50}
\eeq
where $z_i$ ($i=1,2,3$)  are the three roots of the cubic equation
$L_0+S_1 t+S_2 t^2+S_3 t^3=0$. In particular,
\beq
f_0(1)=
\frac{1}{12\eta}\sum_{i=1}^3z_i\frac{L_0-1}{S_1+2S_2z_i+3S_3z_i^2}e^{z_i},
\label{50.2}
\eeq
\beq
f_0'(1)=
\frac{1}{12\eta}\sum_{i=1}^3z_i^2\frac{L_0-1}{S_1+2S_2z_i+3S_3z_i^2}e^{z_i}.
\label{57}
\eeq
One could also evaluate $f_0(0)$ and $f_0'(0)$ from Eq.\ \eqref{50},
but it is easier to do it by a simpler alternative route. Taking
into account the expansion of $F_0(t)$ for large $t$,
\beq
F_0(t)=\frac{L_0-1}{S_3}\frac{t^{-2}}{12\eta}\left(1-\frac{S_2}{S_3}t^{-1}\right)+\mathcal{O}(t^{-4}),
\label{IV.2}
\eeq
one gets
\beq
f_0(r)=\frac{L_0-1}{S_3}\frac{r}{12\eta}\left(1-\frac{S_2}{2S_3}r\right)+\mathcal{O}(r^3).
\label{IV.3}
\eeq
Therefore,
\beq
f_0(0)=0,\quad f_0'(0)=\frac{L_0-1}{12\eta S_3}.
\eeq
 Similarly,
\beq
F_1(t)=-\frac{L_1}{S_3}\frac{t^{-1}}{12\eta}\left[1+\left(\frac{L_0}{L_1}-\frac{S_2}{S_3}\right)t^{-1}\right]+\mathcal{O}(t^{-3}),
\label{IV.8}
\eeq
so that
\beq
f_1(r)=-\frac{L_1}{S_3}\frac{1}{12\eta}\left[1+\left(\frac{L_0}{L_1}-\frac{S_2}{S_3}\right)r\right]+\mathcal{O}(r^{2}),
\label{IV.9}
\eeq
\beq
f_1(0)=-\frac{L_1}{12\eta S_3}, \quad
f_1'(0)=-\frac{{L}_0-{L}_1{{S}_2}/{{S}_3}}{12\eta {S}_3}.
\label{23.2}
\eeq

So far, the parameter $L_0$ remains undetermined. We now fix it  by
imposing the continuity of the cavity function at $r=1$, i.e.,
$y(1^-)=y(1^+)$. {}From Eq.\ (\ref{14}) one has
\beq
r y(r)=\left\{
\begin{array}{ll}
(1-x)^{-1}f_0(r),&0\leq r\leq 1,\\
f_0(r)+f_1(r-1),&1\leq r\leq 2.
\end{array}
\right.
\label{54bis}
\eeq
Therefore,
\beq
y(1^-)=\frac{1}{1-x}f_0(1),\quad y(1^+)=f_0(1)+f_1(0).
\label{22}
\eeq
The condition $y(1^-)=y(1^+)$ yields
\beq
xf_0(1)=(1-x)f_1(0).
\label{23}
\eeq
Insertion of Eqs.\ \eqref{50.2} and \eqref{23.2} into Eq.\
\eqref{23}, together with Eqs.\ \eqref{43bis}--\eqref{46bis},
renders a closed transcendental equation that gives $L_0$ as a
function of $x$ and $\eta$. This closes the proposal \eqref{14}. In
the zero-temperature limit ($x\to 1$), Eq.\ \eqref{23} becomes
$f_0(1)=0$, whose solution is simply $L_0=1$ and then we recover the
analytical PY solution for HSs.

While Eq.\ \eqref{23}  satisfies the condition $y(1^-)=y(1^+)$, it
does not verify $y'(1^-)=y'(1^+)$, except at $T^*=0$. In addition,
it inherits from the PY solution for HSs a poor performance in the
overlapping region, even at low temperatures. Both deficiencies are
remedied by the next stage.

\subsection{Second stage}
In parallel with what we did in Section \ref{sec3} in the case of
the PY solution for HSs [cf.\ Eq.\ \eqref{m3.4}], we propose now in
the case of PSs an improved version of the approximation (\ref{14})
in which only the radial distribution function in the overlapping
region is modified, namely
\beq
g(r)=\frac{e^{Q(r)\Theta(1-r)}}{r}\sum_{n=0}^\infty
f_n(r-n)\Theta(r-n),
\label{24}
\eeq
where $Q(r)$ is a third-degree polynomial with the constraint
$Q(1)=0$ to maintain the continuity of $y(r)$ at $r=1$ obtained in
the first stage. The degree of $Q(r)$ is suggested by the structure
of the exact cavity function to first order in density, Eq.\
(\ref{17.2}), and by the polynomial appearing in the exponential of
Eq.\ \eqref{m3.4}. Thus,  $Q(r)$ can be written as
\beq
Q(r)=(r-1)\left[A+B(r+2)(r-1)+C r(r-1)\right].
\label{IV.1}
\eeq
The parameters $A$, $B$, and $C$ will be determined by imposing the
continuity condition $y'(1^-)=y'(1^+)$, as well as approximate
expressions for the zero-separation values $y(0)$ and $y'(0)$.
According to Eq.\ (\ref{24}), Eq.\ \eqref{54bis} is replaced by
\beq
r y(r)=\left\{
\begin{array}{ll}
(1-x)^{-1}f_0(r)e^{Q(r)},&0\leq r\leq 1,\\
f_0(r)+f_1(r-1),&1\leq r\leq 2.
\end{array}
\right.
\label{54}
\eeq
The condition $y'(1^-)=y'(1^+)$ yields
\beq
A=\frac{x}{f_1(0)}\left[f_1'(0)-\frac{x}{1-x}f_0'(1)\right],
\label{56}
\eeq
where use has been made of Eq.\ (\ref{23}) and $f_0'(1)$, $f_1(0)$,
and $f_1'(0)$ are given by Eqs.\ \eqref{57} and \eqref{23.2}. It
remains to determine the coefficients $B$ and $C$. To that end, note
that Eq.\ (\ref{54}) implies
\beq
y(0)=\frac{e^{2B-A}}{1-x}\frac{L_0-1}{12\eta S_3},
\label{IV.4}
\eeq
\beq
\frac{y'(0)}{y(0)}=A+C-3B-\frac{S_2}{2S_3},
\label{IV.5}
\eeq
where use has been made of Eq.\ (\ref{IV.3}). Therefore, $B$ and $C$
can be expressed in terms of $y(0)$ and $y'(0)$ as follows:
\beq
B=\frac{1}{2}A+\frac{1}{2}\ln\left[(1-x)12\eta\frac{S_3}{L_0-1}y(0)\right],
\label{IV.6}
\eeq
\beq
C=\frac{y'(0)}{y(0)}-A+3B+\frac{S_2}{2S_3}.
\label{IV.7}
\eeq

To close the problem we need to propose approximate expressions for
$y(0)$ and $y'(0)$. Here we take
\beq
\ln y(0)=\frac{11}{2}\eta x^5 y(1)+\ln
\left[y(1)+\frac{11}{2}\eta(x^2-x^5)y^2(1)\right],
\label{IV.10}
\eeq
\beq
\frac{y'(0)}{y(0)}=-6\eta x^2 y(1).
\label{IV.11}
\eeq
Equations (\ref{IV.10}) and (\ref{IV.11}) are exact to first order
in density, as can be seen from Eqs.\ (\ref{II.1a}) and
(\ref{II.1b}). In addition, Eqs.\ (\ref{IV.10}) and (\ref{IV.11})
reduce to Eqs.\ \eqref{11/2} and \eqref{37}, respectively, in the
zero-temperature limit ($x\to 1$). {}From that point of view, Eqs.\
(\ref{IV.10}) and (\ref{IV.11}) can be viewed as approximate
extensions to $x\neq 1$ of the zero-separation theorems. A similar
extension was proposed in the 1D case \cite{MS06}.

To check the reliability of Eqs.\ (\ref{IV.10}) and (\ref{IV.11}),
we have tested them against MC simulations for a wide range of
temperatures and densities. The results are displayed in  Table
\ref{table1}. It can be seen that Eqs.\ (\ref{IV.10}) and
(\ref{IV.11}) compare reasonably well with simulations for all the
states considered.
\begin{table}[htb]
\begin{ruledtabular}
\begin{tabular}{ccccccc}
$T^*$& $\eta$&$y(1)$& $\ln y(0)$&Eq.\ (\protect\ref{IV.10})&$-y'(0)/y(0)$&Eq.\ (\protect\ref{IV.11})\\
\hline
0.1&0.3&2.48&4.85&5.00 &4.27&4.46\\
0.1&0.37&3.16&7.37&7.58 &7.30&7.01\\
0.2&0.3&2.30&4.52& 4.57&5.11&4.08\\
0.3&0.3&1.91&3.53& 3.54&3.65&3.20\\
0.4&0.3&1.67&2.75&2.73 &2.94&2.53\\
0.5&0.3&1.52&2.18&2.14 &2.34&2.05\\
0.5&0.4&1.58&2.85&2.79 &3.48&2.84\\
0.5&0.6&1.59&3.94&3.87 &5.66&4.28\\
1.0&0.4&1.25&1.16&1.10 &1.31&1.20\\
1.0&0.5&1.28&1.38&1.32 &1.67&1.53\\
1.0&0.6&1.29&1.62&1.50 &2.31&1.86\\
1.0&0.8&1.30&2.03&1.84 &2.99&2.49\\
1.7&0.6&1.15&0.72&0.73 &0.87&0.82\\
1.7&0.8&1.16&0.91&0.89 &1.20&1.10\\
1.7&1.2&1.17&1.25&1.16 &1.88&1.67\\
3.0&0.8&1.08&0.37&0.40 &0.44&0.42\\
3.0&1.2&1.09&0.50&0.55 &0.73&0.63\\
3.0&1.8&1.10&0.68&0.73 &0.98&0.95\\
6.0&3.0&1.05 &0.31 &0.39 &0.48 &0.45 \\
20.0&6.0&1.01&0.06&0.09 &0.07&0.09\\
\end{tabular}
\caption{MC simulation data for $y(1)$, $\ln y(0)$, and
$y'(0)/y(0)$. The values obtained from the approximations
(\protect\ref{IV.10}) and (\protect\ref{IV.11}) with the empirical
$y(1)$ are also included.}
\label{table1}
\end{ruledtabular}
\end{table}

This closes the construction of our low-temperature (LT)
approximation. In summary, for any desired density and temperature,
the radial distribution function is given by Eq.\ \eqref{24}, where
$f_0(r)$ is provided by Eq.\ \eqref{50} and $f_n(r)$ with $n\geq 1$
are the inverse Laplace transforms of the functions \eqref{49}. Some
needed specific values are given by Eqs.\ \eqref{50.2}, \eqref{57},
and \eqref{23.2}. The parameters $L_1$, $S_1$, $S_2$, and $S_3$ are
linear functions of $L_0$, as shown by Eqs.\
\eqref{43bis}--\eqref{46bis}, and the latter quantity is the
solution of Eq.\ \eqref{23}. Finally, the polynomial $Q(r)$ is given
by Eq.\ \eqref{IV.1}, where the explicit expressions for the
coefficients $A$, $B$, and $C$ can be found in Eqs.\ \eqref{56},
\eqref{IV.6}, and \eqref{IV.7}, the two latter supplemented by Eqs.\
\eqref{IV.10} and \eqref{IV.11}, where $y(1)$ is given by Eq.\
\eqref{22}. All the expressions are analytical, except Eq.\
\eqref{23}, which must be solved numerically to obtain $L_0$ for
given $\eta$ and $x$. Note that, in order to determine $g(r)$ for
$r<k=\text{integer}$, where in practice $k=3$ is enough, only the
functions $f_n(r)$ with $n\leq k-1$ are needed. They can be
explicitly obtained by application of the residue theorem, as
illustrated in Eq.\ \eqref{50} for $n=0$. Alternatively, one can use
any of the efficient methods discussed by  Abate and Whitt
\cite{AW92} to numerically invert Laplace transforms.

The LT theory for PSs proposed in this Section is within the spirit
of the rational-function approximation that has been applied in the
past to hard-core systems \cite{HYS07}. In fact, as indicated
before, it is straightforward to prove that the LT $g(r)$ given by
Eq.\ \eqref{24} reduces to the PY $g(r)$ in the HS limit ($x\to
1\Rightarrow L_0\to 1$).  The HS limit of the LT $y(r)$ for $r<1$ is
presented in Appendix \ref{appC}. It turns out that the resulting
function, Eq.\ \eqref{26}, gives values that are practically
indistinguishable from those obtained for HSs by  Eqs.\ \eqref{m3.4}
and \eqref{m3.5} implemented with Eq.\ \eqref{11/2}. It is also be
proven in Appendix \ref{appC} that our LT approximation \eqref{24}
for PSs is consistent with the exact form \eqref{17.2} of $y(r)$ to
first order in density.

\section{The high-temperature (HT) approximation \label{sec5}}

The exact asymptotic behavior (\ref{2}) in the limit $T^*\to\infty$
can be used to propose a simple approximation that might be useful
for  high temperatures, in the same spirit as done in the 1D case
\cite{MS06}. The key point  of the approximation is the assumption
that the cavity function $y(r)$ depends on the distance $r$ and the
state variables $x$ and $\eta$ only through the combination $x
w(r)$, where $w(r)$ is the inverse Fourier transform of Eq.\
\eqref{4}, which depends on $x$ and $\eta$ through the product $\eta
x$. This general assumption can be formally expressed as
\beq
y(r)=\mathcal{F}\left(x w(r)\right),
\label{V.1}
\eeq
where the function $\mathcal{F}(z)$ remains to be chosen.
Consistency with Eq.\ (\ref{2}) imposes the constraint
\beq
\mathcal{F}(z)=1+z+\mathcal{O}(z^2).
\label{V.2}
\eeq
We have found that $\mathcal{F}(z)=1+ze^z$ compares with MC
simulations for high temperatures better than other simple
possibilities, such as $\mathcal{F}(z)=1+z$, $\mathcal{F}(z)=e^z$,
or $\mathcal{F}(z)=(1-z)^{-1}$.  Therefore, the high-temperature
(HT) approximation adopted in this paper is
\beq
y(r)=1+x w(r) e^{x w(r)}.
\label{V.8}
\eeq

We have also checked that Eq.\ \eqref{V.8} is superior to the
mean-field approximation $c(r)=f(r)$, where $c(r)$ is the direct
correlation function and $f(r)=-x \Theta(1-r)$ is the Mayer
function. The Ornstein--Zernike relation then yields $g(r)=1+x
w(r)+f(r)$, which implies
\beq
y(r)=1+x w(r)\left[1+\frac{x}{1-x}\Theta(1-r)\right].
\label{V.10}
\eeq
While the mean-field approximation \eqref{V.10} is also consistent
with the exact behavior \eqref{2}, it does not belong in the class
of approximations \eqref{V.1}. In fact, all the approximations of
the form \eqref{V.1} satisfying Eq.\ \eqref{V.2} are consistent with
the exact low-density behavior \eqref{17.2}, while the mean-field
approximation \eqref{V.10} is not.

\section{Comparison with MC simulations\label{sec6}}

\begin{figure}
\includegraphics[width=\columnwidth]{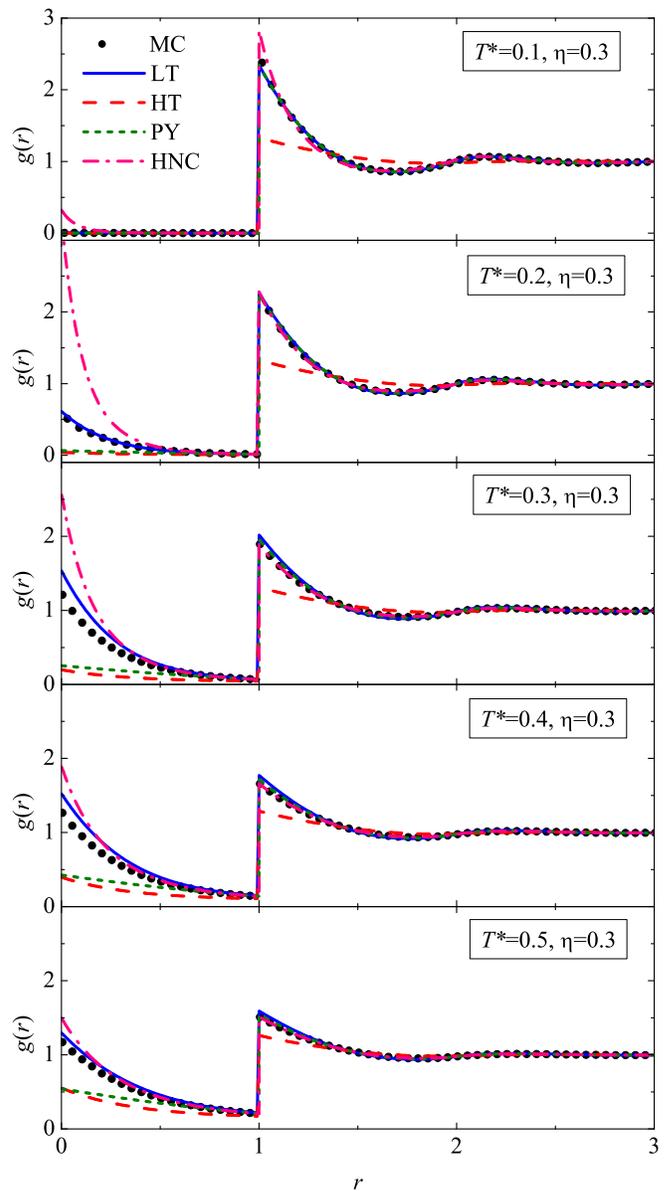}
\caption{(Color online) Plot of the radial distribution  function at
$\eta=0.3$ and for temperatures $T^*=0.1$, $T^*=0.2$, $T^*=0.3$,
$T^*=0.4$, and $T^*=0.5$. The circles are MC simulation data, while
the lines correspond to the LT approximation (---), the HT
approximation (-- -- --), the PY theory (- - -), and the HNC theory
($\text{-- }\cdot\text{ --}$).}
\label{eta0p3BIS}
\end{figure}
\begin{figure}
\includegraphics[width=\columnwidth]{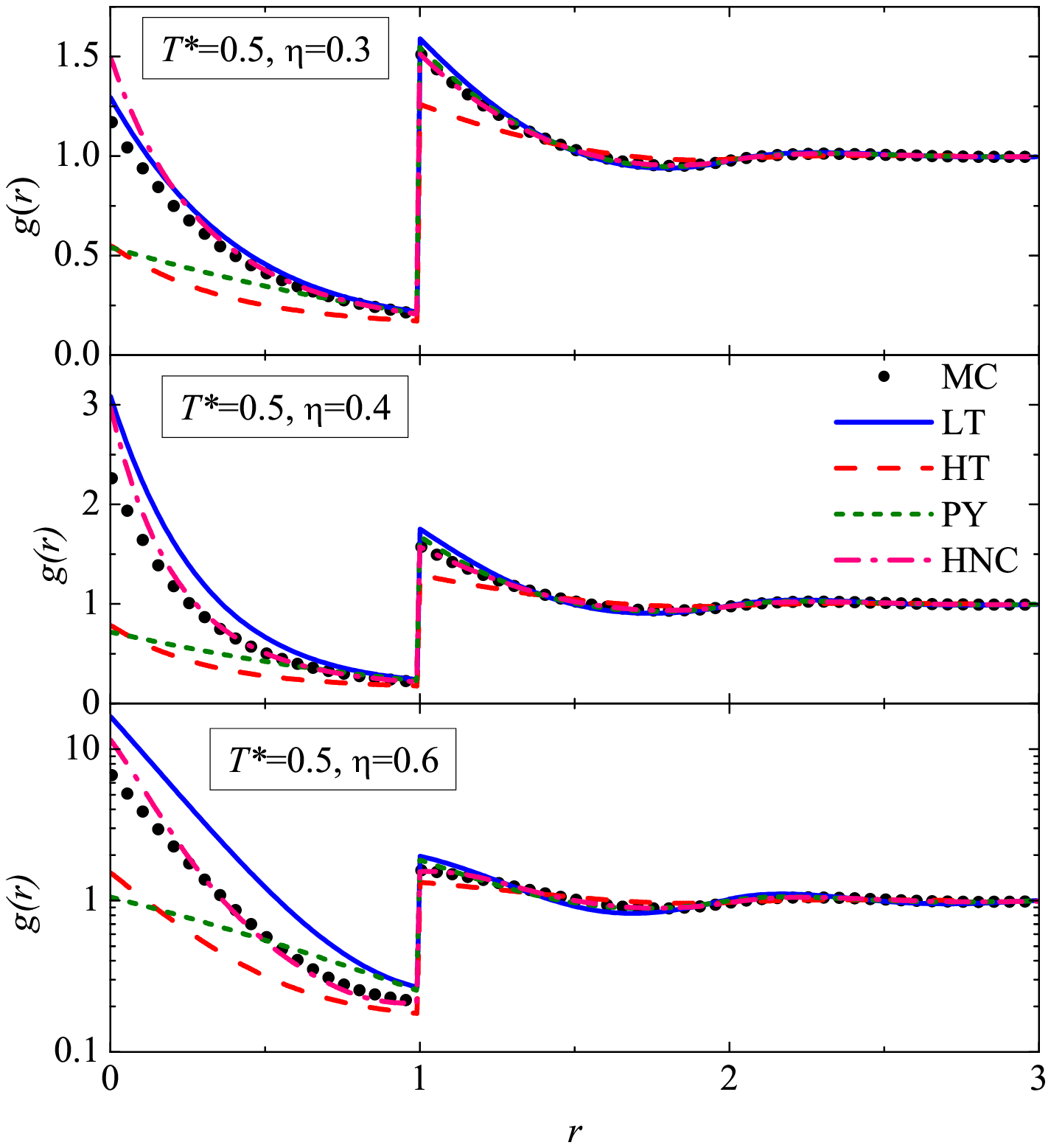}
\caption{(Color online) Plot of the radial distribution  function at
$T^*=0.5$ and for densities $\eta=0.3$, $\eta=0.4$, and $\eta=0.6$.
The circles are MC simulation data, while the lines correspond to
the LT approximation (---), the HT approximation (-- -- --), the PY
theory (- - -), and the HNC theory ($\text{-- }\cdot\text{ --}$).}
\label{T0p5}
\end{figure}
\begin{figure}
\includegraphics[width=\columnwidth]{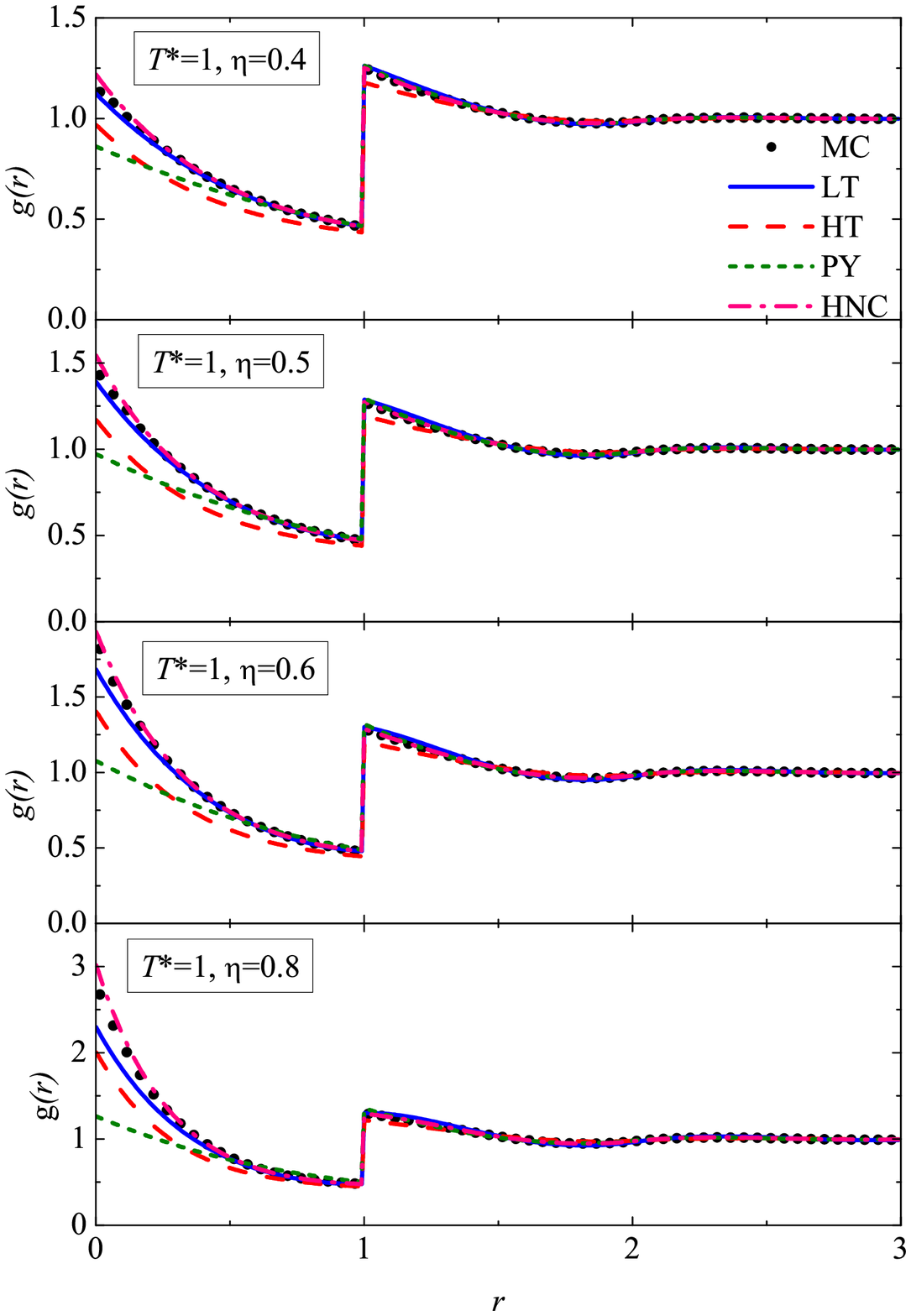}
\caption{(Color online) Plot of the radial distribution  function at
$T^*=1$ and for densities $\eta=0.4$, $\eta=0.5$, $\eta=0.6$, and
$\eta=0.8$. The circles are MC simulation data, while the lines
correspond to the LT approximation (---), the HT approximation (--
-- --), the PY theory (- - -), and the HNC theory ($\text{--
}\cdot\text{ --}$).}
\label{T1}
\end{figure}
\begin{figure}
\includegraphics[width=\columnwidth]{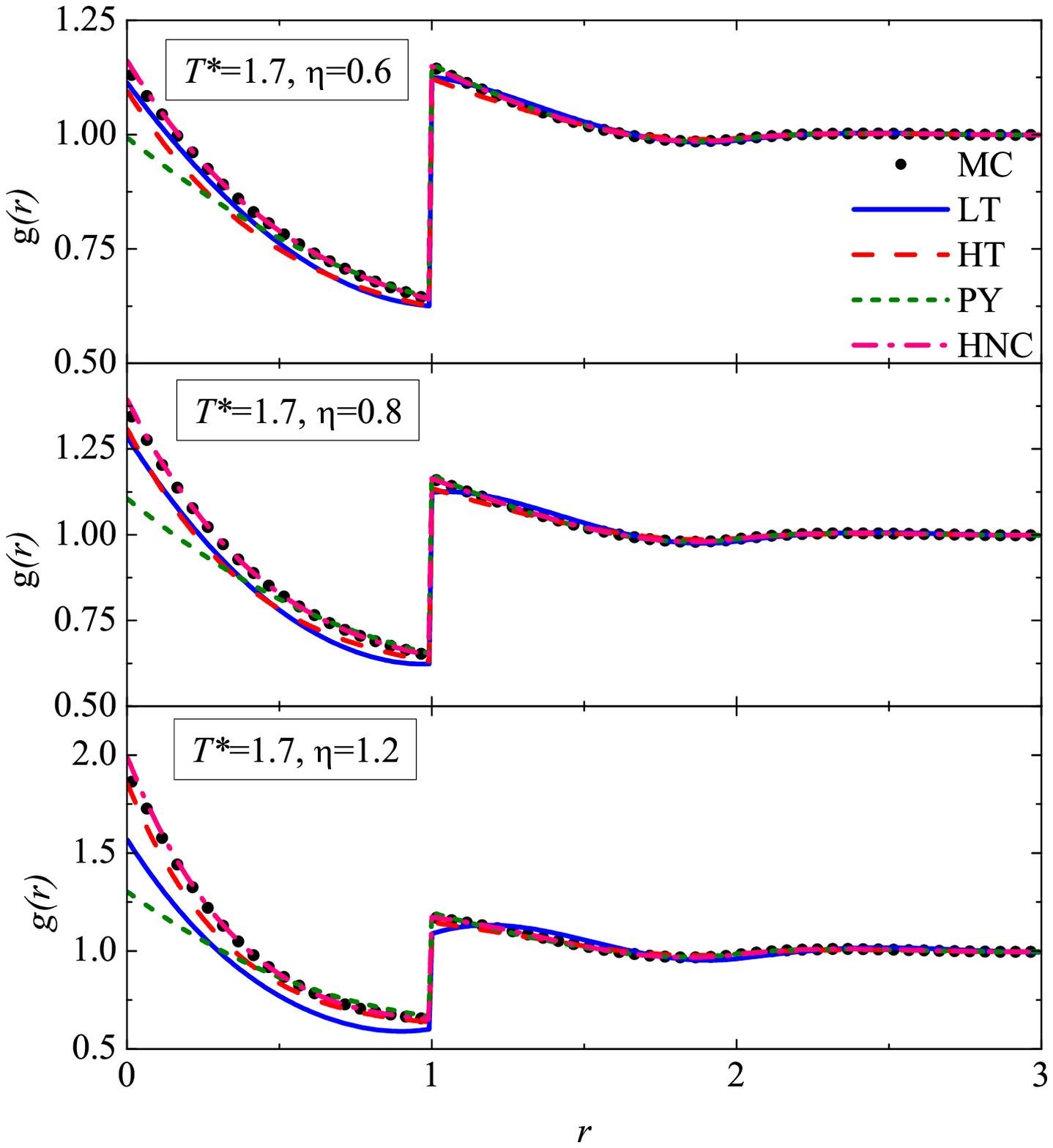}
\caption{(Color online) Plot of the radial distribution  function at
$T^*=1.7$ and for densities $\eta=0.6$, $\eta=0.8$,  and $\eta=1.2$.
The circles are MC simulation data, while the lines correspond to
the LT approximation (---), the HT approximation (-- -- --), the PY
theory (- - -), and the HNC theory ($\text{-- }\cdot\text{ --}$).}
\label{T1p7}
\end{figure}
\begin{figure}
\includegraphics[width=\columnwidth]{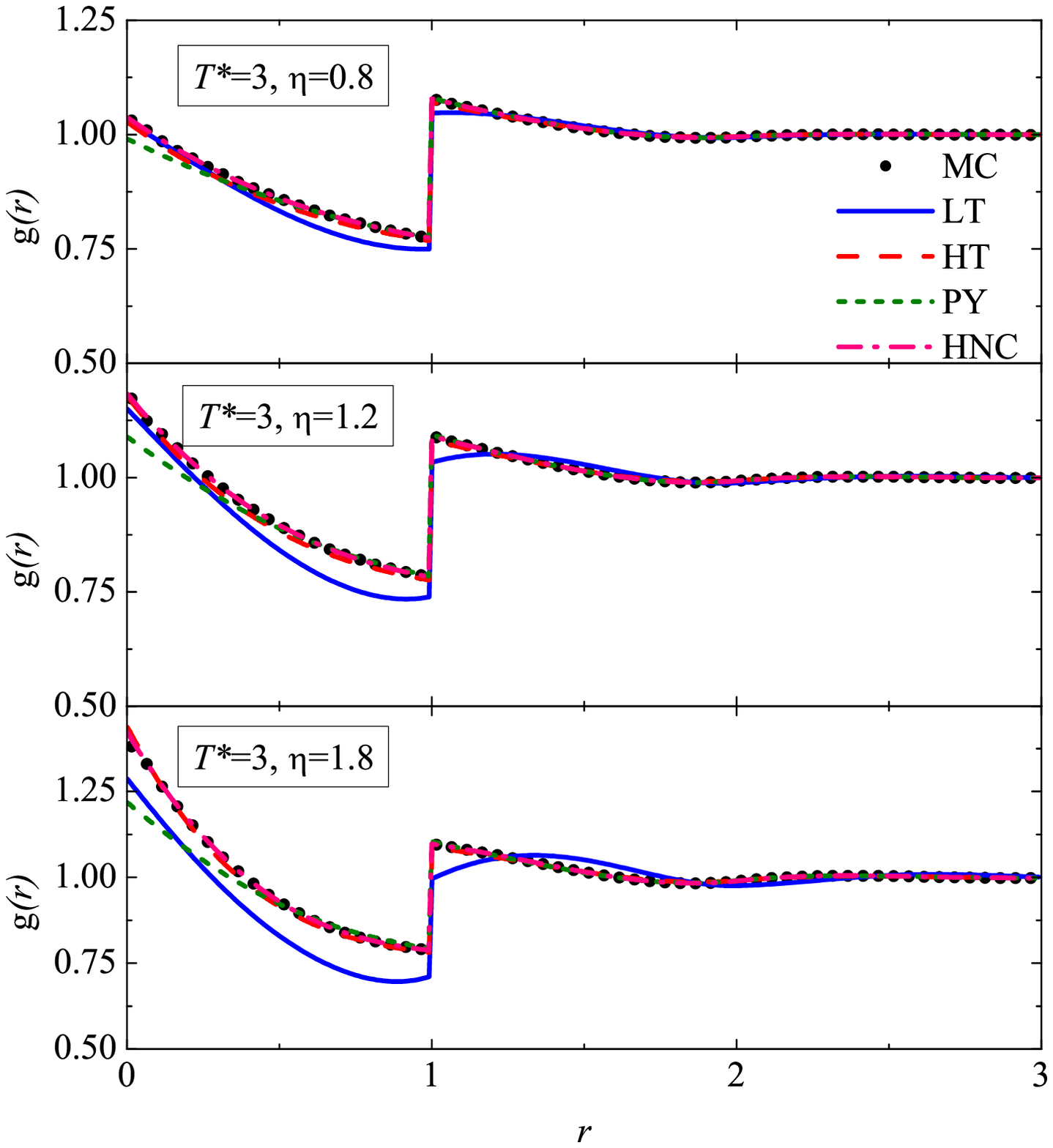}
\caption{(Color online) Plot of the radial distribution  function at
$T^*=3$ and for densities $\eta=0.8$, $\eta=1.2$,  and $\eta=1.8$.
The circles are MC simulation data, while the lines correspond to
the LT approximation (---), the HT approximation (-- -- --), the PY
theory (- - -), and the HNC theory ($\text{-- }\cdot\text{ --}$).}
\label{T3}
\end{figure}
\begin{figure}
\includegraphics[width=\columnwidth]{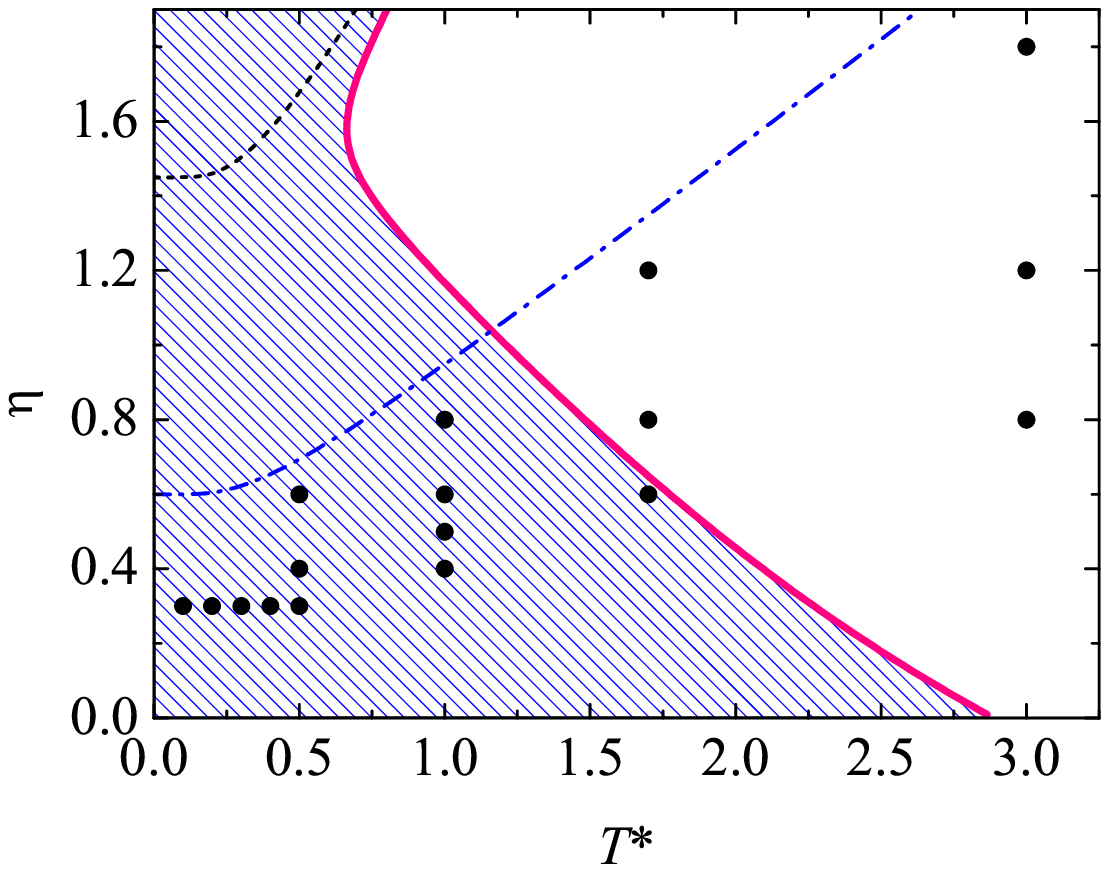}
\caption{(Color online) The circles represent the states considered
in Figs.\ \protect\ref{eta0p3BIS}--\protect\ref{T3}. The thick solid
line represents the locus of points where the LT and HT
approximations predict the same value of the contact quantity
$y(1)=g(1^+)$.  The LT approximation is more accurate in the shaded
region to the left of the curve, while the HT approximation is more
accurate to the right of the curve. The dashed-dotted line is a
rough estimate of the freezing curve, while the dashed line
represents the threshold curve above which the HT theory ceases to
exist.}
\label{locus}
\end{figure}
Here we compare  our two heuristic theories  with MC simulations and
with the results obtained by the numerical solution of  two
classical integral equation theories: the PY and the HNC
approximations.

We first fix the packing fraction at $\eta=0.3$ and increase the
 temperature from $T^*=0.1$ to $T^*=0.5$. The results for  $g(r)$ are shown in
Fig.\ \ref{eta0p3BIS}. We observe that the LT, PY, and HNC theories
practically coincide with the simulation results outside the core
($r>1$), while the HT approximation does it for longer distances
($r\gtrsim 2$). In any case, the interesting region for the PS model
is  the overlapping one ($0\leq r\leq 1$) and it is there where the
largest discrepancies appear.  As expected from the behavior of
$y(r)$ in the zero-temperature limit (see Fig.\ \ref{y_HS_PY}) and
also for low densities \cite{SM07}, the PY theory strongly
underestimates $g(r)$ in the overlapping region. We will see in
subsequent figures that this drawback persists for higher densities
and/or temperatures. As for the HNC theory, it tends to overestimate
$g(r)$ for $r<1$, this effect being dramatic at very low
temperatures (say $T^*=0.1$ and $T^*=0.2$) but rapidly diminishing
its importance as the temperature increases. In fact,  the HNC
theory performs a remarkable good job at $T^*=0.4$ and $T^*=0.5$,
except near the origin.   Concerning the  HT approximation, it is
not surprising that it is so poor, given the range of  temperatures
considered in Fig.\  \ref{eta0p3BIS}. The most noticeable feature is
that our LT approximation, which only requires to solve numerically
Eq.\ \eqref{23}, exhibits a surprisingly good agreement with
simulation for the states considered. Therefore, it achieves its
goal of extending to low temperatures the good role played in the
case of HSs by the (modified) PY theory described in Section
\ref{sec3}. While the temperature $T^*=0.1$ ($1-x\simeq 5\times
10^{-5}$) is so low that the PS fluid with a packing fraction
$\eta=0.3$ behaves practically as the HS system, and so $g(r)\approx
0$ for $r<1$, for temperatures larger than $T^*=0.2$ ($1-x\approx
7\times 10^{-3}$) the number of overlapped pairs becomes observable.
The state $\eta=0.3$ and $T^*=0.2$ was already considered in Refs.\
\cite{LWL98} and \cite{FLL00}, where it was also found that the PY
and HNC theories were highly insufficient in that case.

In Figs.\ \ref{T0p5}, \ref{T1}, \ref{T1p7}, and \ref{T3} we consider
the temperatures $T^*=0.5$, $T^*=1$, $T^*=1.7$, and $T^*=3$,
respectively, and several increasing densities in each case. The
values of the densities are chosen as to make the product $\eta x$
(which is the relevant parameter in the high-temperature limit
\cite{AS04}) ranging from $\eta x\simeq 0.25$ to $\eta x\simeq 0.5$.
The 17 independent states considered in Figs.\
\ref{eta0p3BIS}--\ref{T3} are represented in Fig.\ \ref{locus}.

It is observed in Figs.\ \ref{T0p5}--\ref{T3} that the HNC theory
provides excellent predictions for $T^*\geq 0.5$ at any density. In
contrast, as anticipated before, the PY theory keeps being very
limited at short distances, although it tends to improve as the
temperature increases. With respect to the LT approach, Figs.\
\ref{T0p5}--\ref{T3} confirm that it starts  failing as the
temperature grows. The density also plays an important role since,
at a given temperature, the quality of the LT predictions worsens
with increasing density. In any case, it is remarkable that the LT
theory is rather good for temperatures as high as $T^*=1.7$,
provided that the density is not large  (say $\eta \lesssim 0.6$).
However, the LT theory visibly deteriorates for  higher temperatures
and densities, developing an artificial local minimum at a distance
$r<1$. In contrast, the HT theory, as expected, becomes more
accurate when the temperature and the density increase. In fact at
$T^*=3$ the HT curves practically coincide with the HNC ones.

As shown in Figs.\ \ref{eta0p3BIS}--\ref{T3}, the LT and HT theories
complement each other: the former becomes more accurate as  the
temperature and/or the density decrease, while the opposite happens
for the HT theory. In fact, both heuristic theories  meet and become
practically equivalent at certain intermediate temperatures and
densities, as illustrated by the  top panel of  Fig.\ \ref{T1p7}. In
order to characterize the ``basins of attraction'' of each theory in
the density-temperature plane, let us define the locus of state
points where the contact quantity $y(1)=g(1^+)$ takes the same value
in both approximations. This locus, which is plotted in Fig.\
\ref{locus}, separates the basins of each approximation. As one
departs from this curve, the quality of the corresponding
approximation (either LT or HT) significantly improves, as
illustrated by Fig.\  \ref{eta0p3BIS} in the LT case and by Fig.\
\ref{T3} in the HT case. Figure \ref{locus} also includes the
Kirkwood's instability line $\eta x=1.45$, above which the function
$w(r)$ is not defined. In the high-temperature limit the freezing
transition has been estimated to occur at $\eta x\simeq 0.6$
\cite{AS04}. The extrapolation of this value to any temperature
provides an (admittedly rough) estimate of the freezing curve
(dashed-dotted line in Fig.\ \ref{locus}). Therefore the fluid is
stable approximately below that curve and it is in that region where
the domains of the LT and HT approximations are meaningful.

\section{Summary and concluding remarks\label{sec7}}

In this paper we have investigated the structural properties of the
PS model by computational, numerical, and analytical tools. In the
limit of asymptotically high temperatures  the linear chain diagrams
contributing to the virial expansion of the cavity function
dominate, their sum giving rise to  Eqs.\ (\ref{2}) and (\ref{4}).
 The auxiliary function
$w(r)$ depends on density and temperature {only} through the scaled
quantity $\eta x$, where $x$ is defined by Eq.\ (\ref{3}), and its
series expansion   converges uniformly for $\eta x<\frac{1}{8}$.

In the opposite  zero-temperature limit,  the system becomes an HS
fluid, whose expression for $g(r)$ in the PY approximation is known
exactly in Laplace space. Since the particles are impenetrable in
this limit, $g(r)$ vanishes for $r<1$. However, the cavity function
$y(r)$ remains different from zero, its determination being
important to understand the behavior of $g(r)$ for low (but
non-zero) temperatures, as shown in Fig.\ \ref{y_HS}. We have seen
that the cavity function for $r<1$ predicted by the PY theory for
HSs is extremely poor (cf.\ Fig.\ \ref{y_HS_PY}), this effect being
a precursor of the tendency of the PY theory to dramatically
underestimate the penetrability phenomenon at finite temperatures.
However, the modified version \eqref{m3.4}, implemented with the
consistent zero-separation values \eqref{m3.3} and \eqref{m3.3.2},
represents a significant improvement. Slightly better results are
obtained when Eq.\ \eqref{m3.3} is replaced by Eq.\ \eqref{11/2}.

Based on the explicit expression of $g(r)$ obtained from the PY
approximation in the HS limit, complemented by Eq.\ \eqref{m3.4}, we
have proposed an approximate theory which, by construction, should
be accurate for very low temperatures (LT). This LT approximation is
given by Eq.\ \eqref{24}, where the Laplace transforms of the
functions $f_n(r)$ can be found in Eqs.\ \eqref{48} and \eqref{49},
and the polynomial $Q(r)$ is given by Eq.\ \eqref{IV.1}. The
coefficients appearing in those functions are determined by imposing
consistency conditions, namely the physical requirement  \eqref{17},
the continuity of $y(r)$ and $y'(r)$ at $r=1$, and prescribed
expressions for the zero-separation values $y(0)$ and $y'(0)$. For
the latter two quantities we have proposed approximate extensions to
finite temperatures of the zero-separation theorems holding in the
case of HSs. Those extensions, which are given by Eqs.\
\eqref{IV.10} and \eqref{IV.11}, are reasonably accurate for wide
ranges of temperature and density, as shown in Table \ref{table1}.
All the coefficients are  explicitly expressed in terms of one of
them (chosen to be $L_0$), which is obtained as the solution of a
closed transcendental equation, Eq.\ \eqref{23}, stemming from the
continuity condition $y(1^-)=y(1^+)$.

The LT theory has been complemented by a simpler approximation
constructed by exploiting the exact asymptotic behavior of the
correlation functions for high temperatures.  Among several
possibilities, we have taken Eq.\ \eqref{V.8} as our
high-temperature (HT) approximation, which agrees with simulation
results better than the mean-field approximation \eqref{V.10}.

In order to test the LT and HT theories from moderately low to
moderately high temperatures, we have carried out MC simulations for
the 17 states represented in Fig.\ \ref{locus}. We have also
numerically solved the PY and HNC integral equations for each one of
those states. The results show that the LT theory is the best one
for $T^*\leq 0.4$ and does a very good job for higher temperatures,
provided the density is not too large. It is a reasonable
approximation even at a temperature  $T^*=1.7$ and a density
$\eta=0.6$. However, it rapidly deteriorates for higher temperatures
(for instance, at $T^*=3$). As the LT approximation worsens,  the HT
theory improves and becomes very accurate for $T^*\geq 1.7$. Figure
\ref{locus} shows the regions in the parameter space $\eta$--$T^*$
where each theory is expected to be more reliable

Regarding the classical integral equation theories, we have found
that the HNC approximation is excellent, except for low temperatures
($T^*\leq 0.4$), its performance increasing with  temperature. This
is similar to the situation in 1D \cite{MS06} and agrees with recent
results obtained for the generalized exponential model with $n=4$
\cite{LMGK07}. On the other hand, the PY theory is generally very
poor near $r=0$, especially for low temperatures.

The importance of having  analytical treatments for the structural
properties of fluids at one's disposal cannot be overemphasized.
{}From that point of view, we believe that the LT and HT theories
constructed here can be useful to describe the spatial correlation
functions of the PS model without the burden of numerically solving
integral equations. Both theories complement each other and thus
their combined use  covers satisfactorily well the whole range of
densities and temperatures. Of course, it would be desirable  to
have a unique \emph{analytical} approximation  equally accurate for
both low and high temperatures. However, this does not seem to be an
easy task at all. As a matter of fact, the PS model represents an
interesting benchmark for liquid theory since it encompasses
different regimes: the HS system at very low temperatures and finite
densities, the ideal gas at very high temperatures and finite
densities, and the mean-field system at very high temperatures and
densities.

\acknowledgments

The research of Al.M. has been
 partially supported by the Ministry of Education, Youth, and Sports of
the Czech Republic under Project No.\ LC 512 (Center for
Biomolecules and Complex Molecular Systems) and by the Grant Agency
of the Czech Republic under Projects No.\ 203/06/P432 and No.\
203/05/0725. The research of S.B.Y. and A.S. has been supported by
the Ministerio de Educaci\'on y Ciencia (Spain) through Grant No.\
FIS2007-60977 (partially financed by FEDER funds) and by the Junta
de Extremadura-Consejer\'{\i}a de Infraestructuras y Desarrollo
Tecnol\'ogico.

\appendix

\section{Laplace transform\label{Laplace}}
The Laplace transform of $rg(r)$ is
\beq
 G(t)=\int_0^\infty dr\, e^{-rt}rg(r).
\label{7}
\eeq
The relationship between $G(t)$ and the Laplace transform $H(t)$ of
$rh(r)$, where $h(r)=g(r)-1$ is the total correlation function,  is
\beq
G(t)=\frac{1}{t^2}+H(t).
\label{16}
\eeq
The isothermal compressibility is directly related to $\left.d
H(t)/dt\right|_{t=0}$ and is  finite. Therefore,  $H(t)$ must be
finite at $t=0$, so that the small-$t$ behavior of $G(t)$ must be
\beq
G(t)=\frac{1}{t^2}\left[1+\mathcal{O}(t^2)\right].
\label{17}
\eeq
The large-$t$ behavior of $G(t)$ is also easy to find. Near the
origin, Eq.\ \eqref{2.1} implies that $g(r)=(1-x)\left[ y(0)+
y'(0)r+\mathcal{O}(r^2)\right]$, so that
\beq
G(t)=\frac{1-x}{t^2}\left[y(0)+2
y'(0)t^{-1}+\mathcal{O}(t^{-2})\right].
\label{17bis}
\eeq

Now we define an auxiliary function $P(t)$ by
\beq
G(t)=\frac{t}{12\eta}\frac{P(t)}{1+ P(t)}.
\label{9}
\eeq
Thus, Eqs.\ \eqref{17} and \eqref{17bis} are equivalent to
\beq
P(t)=-1-\frac{1}{12\eta}t^3+\mathcal{O}(t^5),
\label{17.2.1}
\eeq
\beq
P(t)=12\eta\frac{1-x}{t^3}\left[y(0)+2
y'(0)t^{-1}+\mathcal{O}(t^{-2})\right],
\label{17bis.2}
\eeq
respectively.

\section{Virial expansion of the function $w(r)$\label{app0}}
 The inverse Fourier transform of $\widetilde{w}(k)$ is
\beq
w(r)=\frac{1}{2\pi^2 r}\int_0^\infty dk\, k\sin(kr)
\widetilde{w}(k).
\label{0.3}
\eeq
Note now that Eq.\ \eqref{4} can be rewritten as
\beq
\widetilde{w}(k)=\frac{\pi}{6\eta
x}\frac{\widetilde{F}^2(k)}{1-\widetilde{F}(k)}= \frac{\pi}{6\eta
x}\sum_{n=2}^\infty \widetilde{F}^n(k),
\label{0.1}
\eeq
where we have called
\beq
\widetilde{F}(k)\equiv 24\eta x \frac{k\cos k-\sin k}{k^3}.
\label{0.2}
\eeq
Therefore, Eq.\ \eqref{0.3} becomes
\beq
w(r)=\sum_{n=2}^\infty (\eta x)^{n-1} w_n(r)
\label{0.4}
\eeq
with
\beq
w_n(r)=\frac{ 24^{n}}{12\pi r}\int_{0}^\infty dk \, k
\sin(kr)\left(\frac{k\cos k-\sin k}{k^3}\right)^n.
\label{0.5}
\eeq
Application of the residue theorem gives, after some standard but
lengthy manipulations, the result
\beq
w_n(r)=\frac{1}{r}\sum_{m=0}^{[(n-1)/2]}(n-2m-r)^{2(n-1)}\mathcal{P}_n^{(m)}(r)\Theta(n-2m-r).
\label{0.6}
\eeq
Here, $[k]$ denotes the integer part of $k$ and
$\mathcal{P}_n^{(m)}(r)$ is the polynomial of degree $n$ given by
\beq
\mathcal{P}_n^{(m)}(r)=\sum_{p=0}^n c_{np}^{(m)}(n-2m-r)^{n-p},
\label{0.7}
\eeq
with the coefficients
\beqa
c_{np}^{(m)}&=&(-1)^m\frac{12^{n-1}
n!}{(3n-2)!}\binom{3n-2}{p}\sum_{q=0}^p \binom{p}{q}\nn
&&\times\frac{(-1)^q}{(m-p+q)!(n-m-q)!}.
\label{0.8}
\eeqa
{}From Eq.\ \eqref{0.6} it is possible to check the relationship
\beq
\frac{w_n(1)}{n}+\frac{w_{n+1}(0)}{8(n+1)}=0,\quad n\geq 2,
\eeq
which allows one to prove the thermodynamic consistency between the
virial and energy routes to the equation of state \cite{AS04}. The
first few functions $w_n(r)$ are
\beq
w_2(r)=\frac{1}{2}(r-2)^2(r+4)\Theta(2-r),
\eeq
\beqa
w_3(r)&=&\frac{3(r-1)^4}{35r}(r^3+4r^2-53 r-162)\Theta(1-r)\nn
&&-\frac{(r-3)^4}{35r}(r^3+12r^2+27 r-6)\Theta(3-r), \nn &&
\eeqa
\beqa
w_4(r)&=&-\frac{(r-2)^6}{525r}(r^4+12r^3-96r^2-1232 r-2304)\nn
&&\times \Theta(2-r)+\frac{(r-4)^6}{2100r}(r^4+24r^3+156r^2\nn
&&+224 r-144)\Theta(4-r).
\eeqa

The series \eqref{0.4} converges for $\eta x<\frac{1}{8}$ because
$\widetilde{w}(k)$ is singular at $k=0$ when $\eta x$ takes the
negative value $\eta x=-\frac{1}{8}$. By an adequate reordering of
terms, Eq.\ \eqref{0.4} can be rewritten as
\beqa
w(r)&=&-\Theta(1-r)+r^{-1}\sum_{n=1}^\infty \Theta(n-r)\nn
&&\times\sum_{m=0}^\infty\left[\eta
x(n-r)^2\right]^{n+2m-1}\mathcal{P}_{n+2m}^{(m)}(r).
\eeqa
Therefore, $w(r)$ has a fourth-order discontinuity at $r=1$ and a
discontinuity of order $2(n-1)$ at $r=n\geq 2$. For instance,
\beq
w''''(1^+)-w''''(1^-)=432 (\eta x)^2,
\eeq
\beq
w''(2^+)-w''(2^-)=-6 \eta x,
\eeq
\beq
w''''(3^+)-w''''(3^-)=48 (\eta x)^2.
\eeq

\section{Some consequences of  the LT theory\label{appC}}

\subsection{Zero-temperature limit of the cavity function inside the
core\label{appC1}}

According to Eq.\  \eqref{54}, the cavity function inside the core
is
\beq
y(r)=(1-x)^{-1}\frac{f_0(r)}{r} e^{Q(r)}.
\label{C1}
\eeq
Now we will take the limit $\epsilon\equiv 1-x=e^{-1/T^*}\to 0$.
 Let us first write
$L_0=1-K\epsilon+\mathcal{O}(\epsilon^2)$. Thus, Eq.\ (\ref{50})
becomes $f_0(r)\to K\epsilon \bar{f}_0(r)$, where
\beq
\bar{f}_0(r)= -\frac{1}{12\eta}\sum_{i=1}^3\frac{z_i
e^{z_ir}}{S_1+2S_2 z_i+3S_3 z_i^2},
\label{62}
\eeq
the parameters $L_1$, $S_1$, $S_2$, and $S_3$ being now given by
Eqs.\ \eqref{43}--\eqref{46}. Near the origin, following steps
similar to those of Eq.\ \eqref{IV.3}, one obtains
\beq
\bar{f}_0(r)=\frac{1+2\eta}{(1-\eta)^2}r\left[1-\frac{3\eta}{1-\eta}r+\frac{3\eta^2}{(1-\eta)^2}r^2\right]+\mathcal{O}(r^4).
\label{66}
\eeq
According to Eq.\ (\ref{23}), the
coefficient $K$ is explicitly given by
\beq
K=\frac{y_\py(1)}{\bar{f}_0(1)}.
\label{63}
\eeq
Therefore,  Eq.\ (\ref{C1}) becomes
\beq
y(r)=K\frac{\bar{f}_0(r)}{r}e^{Q(r)},
\label{26}
\eeq
where the coefficients $A$, $B$, and $C$ of the polynomial $Q(r)$
reduce to [cf.\ Eqs.\ \eqref{IV.1}, \eqref{56},
\eqref{IV.6}--\eqref{IV.11}]
\beq
A=1+\frac{y_\py'(1)}{y_\py(1)}-\frac{\bar{f}_0'(1)}{\bar{f}_0(1)},
\label{64}
\eeq
\beq
B=\frac{1}{2}A+\frac{1}{2}\ln\left[\frac{y_\mpy(0)(1-\eta)^2}{(1+2\eta)K}\right],
\label{65}
\eeq
\beq
C=\frac{y_\mpy'(0)}{y_\mpy(0)}-A+3B+\frac{3\eta}{1-\eta}.
\label{65.2}
\eeq

\subsection{Low-density limit
\label{appB}} To first
order in density, the parameter $L_0$ has the form
$L_0=L_{00}+L_{01}\eta+\mathcal{O}(\eta^2)$. The coefficients
$L_{00}$ and $L_{01}$ are  determined by expanding $F_0(t)$ and
$F_1(t)$ to first order in $\eta$ and imposing the condition
$y(1^-)=y(1^+)$ to that order. This yields, after some algebra,
\beq
L_0=x+\frac{1}{2}\eta x(1-x)(3-5x)+\mathcal{O}(\eta^2).
\label{52b}
\eeq
The functions $f_0(r)$ and $f_1(r)$ are given by
\beq
f_0(r)=(1-x)r\left[1+\frac{1}{2}\eta
x(5+5x-6r+r^3)+\mathcal{O}(\eta^2)\right],
\eeq
\beq
f_0(r)+f_1(r-1)=r\left[1+\frac{1}{2}\eta
x^2(r-2)^2(r+4)+\mathcal{O}(\eta^2)\right].
\eeq
Finally, use of Eqs.\ \eqref{56}, \eqref{IV.6}, and \eqref{IV.7}
gives
\beq
A=\frac{3}{2}\eta x(1-3x)+\mathcal{O}(\eta^2),
\eeq
\beq
B=-\frac{1}{2}\eta x(1-x)+\mathcal{O}(\eta^2),\quad
C=\mathcal{O}(\eta^2)
\eeq
It is then straightforward to check that  Eq.\ \eqref{24} reduces to
Eq.\ \eqref{17.2}.


\begin{thebibliography}{99}
\bibitem{note} In the MC simulations presented in this paper
{we have employed the conventional Metropolis algorithm on an $NVT$
ensemble with periodic boundary conditions. Typically, $N=10^4$
particles are used. In each realization, the correlation functions
are measured every 100 MC sweeps and this cycle is repeated $2\times
10^4$ times. The time-averaged quantities are further averaged over
10 independent realizations.}
\end{thebibliography}
\end{document}